\documentclass[journal, 12pt, onecolumn]{IEEEtran}

\usepackage{url}
\usepackage{blindtext}
\usepackage{subcaption}
\usepackage{multirow}
\usepackage{graphicx}
\usepackage{cite}
\usepackage{amsfonts}

\usepackage{graphicx}
\usepackage{amsmath,amssymb,amsfonts}
\usepackage{amsthm}
\usepackage{mathrsfs}
\usepackage{xcolor}
\usepackage{textcomp}
\usepackage{manyfoot}
\usepackage{listings}

\usepackage{subcaption}
\usepackage{amsmath}

\usepackage{multicol}
\usepackage{multirow}
\usepackage{algpseudocode}
\usepackage{algorithm}

\begin{document}

\title{Fast Adaptive Regression-based Model Predictive Control}

\author{Eslam~Mostafa,
        Hussein~A.~Aly,
        Ahmed~Elliethy,
\thanks{Authors are with the Department of Electrical and Computer Engineering, Military~Technical~College, Cairo 14627, Egypt (e-mails: eslammostafafawzy@mtc.eg.edu, haly@ieee.org, a.s.elliethy@mtc.eg.edu).}
\thanks{This    paper    has    supplementary    downloadable    material    available  provided  by  the  authors.  The  material  consists  of a supplementary document.}
\thanks{Color  versions  of  one  or  more  of  the  figures  in  this  paper  are  available in the electronic version of this manuscript.}}


\maketitle

\begin{abstract}

Model predictive control (MPC) is an optimal control method that predicts the future states of the system being controlled and estimates the optimal control inputs that drive the predicted states to the required reference. The computations of the MPC are performed at pre-determined sample instances over a finite time horizon. The number of sample instances and the horizon length determine the performance of the MPC and its computational cost. A long horizon with a large sample count allows the MPC to better estimate the inputs when the states have rapid changes over time, which results in better performance but at the expense of high computational cost. However, this long horizon is not always necessary, especially for slowly-varying states. In this case, a short horizon with less sample count is preferable as the same MPC performance can be obtained but at a fraction of the computational cost. In this paper, we propose an adaptive regression-based MPC that predicts the best minimum horizon length and the sample count from several features extracted from the time-varying changes of the states. The proposed technique builds a synthetic dataset using the system model and utilizes the dataset to train a support vector regressor that performs the prediction. The proposed technique is experimentally compared with several state-of-the-art techniques on both linear and non-linear models. The proposed technique shows a superior reduction in computational time with a reduction of about 35-65\% compared with the other techniques without introducing a noticeable loss in performance.

\end{abstract}


\newcommand{\grx}{\chi}
\newcommand{\gry}{\zeta}
\newcommand{\rgrx}{^r\grx}
\newcommand{\rgry}{^r\gry}
\newcommand{\betaB}{{\boldsymbol\beta}}
\newcommand{\omegaB}{{\boldsymbol\omega}}
\newcommand{\gammaB}{{\boldsymbol\gamma}}
\newcommand{\alphaB}{{\boldsymbol\alpha}}
\newcommand{\segCI}{{\alphaB_c^i}}
\newcommand{\featureCI}{{\betaB_c^i}}
\newcommand{\featureC}{{\betaB_c}}
\newcommand{\refx}{\mathbf{r}}
\newcommand{\refxi}{r}
\newcommand{\refu}{\mathbf{v}}
\newcommand{\refui}{v}
\newcommand{\Figs}{./Figs}

%
\IEEEpeerreviewmaketitle

\section{Introduction}\label{sec1}

Model predictive control (MPC)~\cite{garcia1989MPCbook} is an advanced control method that has been widely used in many applications\cite{hrovat2012MPCAutomotive,bakovsova2014MPCApp,Wang2018MPCapp1,song2020adaptive,eslam2022mpcApp4}. Given a discrete system with its states defined at specific sampling instances, the MPC utilizes the mathematical model of the system to predict its future states at each sample instant over a finite future horizon time. The predicted states along with a set of given system constraints are used to formulate an optimization problem that is solved \textit{online} in every control cycle to estimate the optimal control inputs at each sample instant over the horizon. Only the estimated input at the first sample instant is applied to the system, and the MPC repeats the same process for all subsequent control cycles. For a long horizon with a large number of sample instances, the predicted behavior of the system becomes more intimate to the required reference \cite{maciejowski2002MPCBook1}, i.e., better control performance. However, for such a long horizon, the MPC easily may not finalize the computations involved in solving the online optimization problem within the control cycle, and this results in a lagging control input to the system in this case.

To avoid the aforementioned lagging control input problem, several studies in the literature focus on speeding up the MPC by reducing its online computations. For example, several techniques try to minimize the number of sample instances within the horizon. In~\cite{khan2013parm,muehlebach2019parm,hyatt2020parameterized}, the input and state trajectories, that represent their variations over time, is parametrized with some basis functions. The parameterization of the trajectories reduces the degrees of freedom of the online optimization problem by calculating the control inputs only at specific sampling instances in the horizon while evaluating the rest using the parametrized version. In~\cite{funke2016dualMPC,ogren2010dualMPC,kim2013dualMPC}, non-uniformly spaced sample instances are used  such that smaller intervals between sample instances are used with the near future of the horizon while larger ones are used with the distant future. These techniques use a fixed horizon but keep the number of sample instances relatively small. However, using a fixed horizon is not optimal in all control scenarios. As shown in~\cite{rezaei2015AdaptiveHorizon}, a short horizon is enough for controlling a vehicle on highways where speed fluctuations are not fast, while a longer horizon is needed when driving in a city due to higher fluctuations in speed and environmental variables. Following this, the technique in~\cite{kim2021dual} adaptively changes the horizon according to the curvature value for the state trajectories. However, the technique uses a  heuristic rule to determine the horizon length, which results in a sub-optimal horizon. In~\cite{shekhar2012Roubstadaptivehorizon}, a variable horizon MPC is achieved by defining several fixed-horizon optimization problems with different horizon lengths. The overall complexity is relatively reduced by utilizing the time-varying move blocking technique~\cite{Cagienard:2004:MoveBlocking} which fixes adjacent in-time decision variables of the optimization problem or its derivatives to be constant over several control cycles, and thus it results in sub-optimal control performance. In~\cite{scokaert1998minmaxAdaptiveHorizon}, the horizon length is added as an extra degree of freedom to the MPC formulation, and thus it incurs additional computations in the control cycle. The technique in~\cite{krener2018stableAdaptiveHorizon} deals with systems with non-linear dynamics and incrementally adjusts the horizon length to its minimum possible value that guarantees stabilization. This incremental adjusting for the horizon is practically not suitable for fast applications.

Another line of research is the so-called explicit MPC~\cite{bemporad2000explicit,bemporad2002explicit,alessio2009explicit}, which pre-computes the optimal control inputs \textit{offline} as a function of the current state and reference states. Thus, the online optimization reduces to a simple search within the pre-computed values. The major drawback here is the searching time for the optimal solution which quickly increases when the number of states, horizon length, or the number of control inputs increases. Thus the explicit MPC can only be applied to small problems. This drawback is partially mitigated in~\cite{pannocchia2007explictAPP} which uses a partial enumeration technique that offline computes and stores in a table the optimal control inputs for only frequently occurring constraint sets. The table is searched online for the best control and updated to incorporate new constraint sets as the control progresses in time.

Recently, machine learning techniques are employed for horizon prediction. In~\cite{bohn2021reinforcement, bohn:2021:RLoptimization}, a novel technique is proposed that uses reinforcement learning (RL)~\cite{sutton2018reinforcementBook} to predict the optimal horizon length using a policy function of the current state. The policy of the RL is modeled as a neural network (NN)~\cite{suzuki2013neuralBook} which is trained using the data collected during the operation of the system being controlled. The training of the NN is performed online within each control cycle, which adds more computations within the cycle. Another technique in~\cite{gardezi2018neuralNetwork} trains a NN offline that is used to predict the optimal horizon at run-time. However, these techniques perform the prediction solely based on the \textit{instantaneous} state of the system, without any consideration for the future values of the states. Moreover, the techniques do not employ any feature engineering for training the NN, and thus it easily over-fits the training data. Additionally, the technique in~\cite{bohn2021reinforcement} predicts the horizon length only while the technique in~\cite{gardezi2018neuralNetwork} uses the move-blocking strategy~\cite{Cagienard:2004:MoveBlocking} that fixes the ratio between the number of sample instances and the time span of the predicted horizon. Therefore, the predictions performed by these techniques are not optimal in general.

In this paper, our goal is to accurately estimate both the best minimum horizon length and the number of samples without introducing a noticeable loss in the performance of the MPC. To this goal, we first propose a mathematical formulation that relates the horizon length and the sample count with the performance of the MPC, then we propose an efficient solution for it. Specifically, we propose an adaptive regression-based MPC (ARMPC) that predicts the best minimum horizon length and the sample count according to the current and future variations exhibited in the reference states of the model. To train the regression, we build a dataset by extracting several features that capture the variations of the reference trajectories of the model over future time with the associated best horizon and sample count in each situation. In run time, we extract the same set of features which presented to the regressor to predict the best minimum horizon length and the sample count. 

Compared with previous techniques, our proposed ARMPC has several advantages. First, it estimates \textit{both} the best horizon length and the best number of samples on the horizon, and this allows the proposed technique to provide more reduction in computations. Second, our technique does not rely on raw values of the reference states but employs feature engineering to extract several distinctive features from the reference trajectories, and thus it avoids over-fitting in the learning phase. Finally, these features are extracted not from the instantaneous values of the states but over a future span of the horizon, which allows a more accurate estimate for the horizon and the sample count. These advantages are reflected in our experimental results where we compared the proposed ARMPC with three different state-of-the-art techniques on both linear and non-linear models. The proposed ARMPC shows a superior reduction in computational time with a reduction of about 35-65\% compared with the other techniques without introducing a noticeable loss in performance. The source code of the proposed technique will be available\footnote{https://github.com/ahmed-elliethy/fast-regression-mpc} online upon acceptance.

The remainder of this paper is organized as follows. Section \ref{sec:MPCBackground} presents a background for the MPC and briefly outlines its parametrization. Section \ref{sec:Motivationandproblemstatement} presents motivation examples and formulates our problem statement. Section~\ref{sec:ProposedadaptiveregressionbasedMPC} introduces the proposed adaptive regression based MPC. Section~\ref{sec:Experimentalresults} describes the experimental setup and discusses the experimental results that evaluate our proposed technique. Section \ref{sec:Conclusion} summarizes our conclusion and presents our future work.

\section{MPC Background}
\label{sec:MPCBackground}

A continuous linear time-invariant system can be described in state space form as~\cite{hyatt2020parameterized} 
\begin{equation}
\dot{\mathbf{x}}(t)=\mathbf{A} \mathbf{x}(t)+\mathbf{B}\mathbf{u}(t),
\label{eq:clti}
\end{equation}
where $t \in \mathbb{R}$ is a continuous time instant, $\mathbf{x}(t) \in \mathbb{R}^{m \times 1}$ and $\mathbf{u}(t) \in \mathbb{R}^{n \times 1}$ represent a vector of states and a vector of system inputs at time $t$, respectively, $\mathbf{A} \in \mathbb{R} ^ {m \times m}$ is the state matrix, and $\mathbf{B} \in \mathbb{R} ^ {m \times n}$ is the input matrix. The system can be discretized using any of the discretization methods (such as the Euler method) by
\begin{equation*}
\dot{\mathbf{x}}_k = \frac{\mathbf{x}_{k+1} - \mathbf{x}_{k}}{t_s},
\end{equation*}
where $k \in \mathbb{Z}$ is a discrete-time instant, $t_s$ is the sampling time, and $\mathbf{x}_{k} = [ x^1_{k}, \dots, x^m_{k} ] ^T$ represents the vector of states at $k$. With the discretization, \eqref{eq:clti} can be written as
\begin{equation}
\mathbf{x}_{k+1}=\mathbf{A}_d \mathbf{x}_{k}+ \mathbf{B}_d \mathbf{u}_{k},
\label{eqn:statespaceform}
\end{equation}
where $\mathbf{u}_{k} = [ u^1_{k}, \dots, u^n_{k} ] ^T$ represents the vector of system inputs at $k$, $\mathbf{A}_d = \mathbf{I} + t_s \mathbf{A}$, $\mathbf{B}_d = t_s \mathbf{B}$, and $\mathbf{I}$ is the identity matrix.

The goal of MPC is to estimate the optimal vector of system inputs for a fixed horizon length of $\mathcal{T}$ time steps in control cycles. More clearly, in the $c^{\text{th}}$ control cycle, the MPC estimates the optimal vector of system inputs from the time instant $c$ to $c+\mathcal{T}-1$. Then only the first control input (at the time instant $c$) is applied to the system and new system states are predicted. In the next cycle, the MPC estimates the system inputs using the newly predicted states, then only the first input is applied again to the system. This loop will be repeated to find the optimum control input in every control cycle.

To do so, in the $c^{\text{th}}$ control cycle, the MPC first expresses the system states as a function of only the inputs and the initial state $\mathbf{x}_{c}$ in the cycle as
\begin{equation}
\begin{aligned}
\left[\begin{array}{c}
\mathbf{x}_{c+1} \\
\mathbf{x}_{c+2} \\
\vdots \\
\mathbf{x}_{c+\mathcal{T}}
\end{array}\right]
&= 
\left[\begin{array}{c}
\mathbf{A}_{d} \mathbf{x}_{c}+\mathbf{B}_{d} \mathbf{u}_{c}  \\
\mathbf{A}_{d} \mathbf{x}_{c+1}+\mathbf{B}_{d} \mathbf{u}_{c+1}  \\
\vdots \\
\mathbf{A}_{d} \mathbf{x}_{c+\mathcal{T}-1}+\mathbf{B}_{d} \mathbf{u}_{\mathcal{T}-1} \end{array}\right]\\
& = \mathbf{S}_c \mathbf{z}_c+\mathbf{n}_c,  
\end{aligned}
\label{eqn:states_vector}
\end{equation}
where
\begin{equation*}
\begin{aligned}
\mathbf{S}_c & = \left[\begin{array}{ccccc}
\mathbf{B}_{d} & \mathbf{0}  &\mathbf{0}& \cdots & \mathbf{0} \\
\mathbf{A}_{d} \mathbf{B}_{d} & \mathbf{B}_{d} &\mathbf{0} & \cdots & \mathbf{0} \\
& \vdots & & \\
\mathbf{A}_{d}^{\mathcal{T}-1} \mathbf{B}_{d} & \mathbf{A}_{d}^{\mathcal{T}-2} \mathbf{B}_{d} &  \mathbf{A}_{d}^{\mathcal{T}-3} \mathbf{B}_{d} & \ldots  & \mathbf{B}_{d} 
\end{array}\right],
\mathbf{n}_c = 
\left[\begin{array}{l}
\mathbf{A}_{d} \mathbf{x}_{c}  \\
\mathbf{A}_{d}^{2} \mathbf{x}_{c}   \\
\vdots \\
\mathbf{A}_{d}^{\mathcal{T}} \mathbf{x}_{c}
\end{array}\right],
\end{aligned}
\end{equation*}
and $ \mathbf{z}_c \in \mathbb{R}^{n\mathcal{T} \times 1} =\left[\mathbf{u}_{c}^{T},\mathbf{u}_{c+1}^{T}, \ldots, \mathbf{u}_{c+\mathcal{T}-1}^{T}\right]^{T}$. Then, the MPC estimates  the optimal vector of system inputs $\mathbf{z}_c^{*}$ that minimizes a cost function $J$ in every control cycle over $\mathcal{T}$ time steps as~\cite{hyatt2020parameterized}
\begin{equation}
    \mathbf{z}_c^{*} = \arg \min_{\mathbf{z}_c} J(\mathbf{z}_c, \mathcal{T}),
    \label{eqn:MPCMathmaticalForm}
\end{equation}
where
\begin{equation}
\begin{aligned}
&J(\mathbf{z}_c, \mathcal{T})=\sum_{k=c}^{c+\mathcal{T}-1}\left(\left[\refx_k-\mathbf{x}_{k}\right]^{T} \mathbf{Q}\left[\refx_k-\mathbf{x}_{k}\right]\right. \left.+\left[\refu_k-\mathbf{u }_{k}\right]^{T} \mathbf{R}\left[\refu_k-\mathbf{u }_{k}\right]\right), \\
&\text { s.t. } \\
&\mathbf{x}_{k+1}=\mathbf{A}_{d} \mathbf{x}_{k}+\mathbf{B}_d \mathbf{u}_{k} \quad \forall k=c, \ldots, c+\mathcal{T}-1,\\
&\mathbf{x_{\text{min} } } \leq \mathbf{x}_{k} \leq \mathbf{x_{\text{max} } } \quad \forall k=c, \ldots, c+\mathcal{T}-1, \\
&\mathbf{u_{\text{min}} } \leq \mathbf{u }_{k} \leq \mathbf{u_{\text{max}} } \quad \forall k=c, \ldots, c+\mathcal{T}-1, \\
&\Delta \mathbf{u_{\text{min}} } \leq \Delta \mathbf{u }_{k} \leq \Delta \mathbf{u_{\text{max}} } \quad \forall k=c, \ldots, c+\mathcal{T}-1. \\
\end{aligned}
\label{eqn:MPCCost}
\end{equation}
In the above cost function, $\refx_{k} = [\refxi^1_k, \dots, \refxi^{m}_{k}] ^{T}$ and $\refu_{k} = [\refui^1_k, \dots, \refui^{n}_{k}] ^{T}$ are the reference states and reference inputs at $k$, respectively. $\mathbf{x_{\text{min} } }$, and $\mathbf{x_{\text{max}}}$ are the minimum and maximum bounds of the states, respectively. Similarly, $\mathbf{u}_{\text{min}}$ and $\mathbf{u_{\text{max}}}$ are the minimum and maximum bounds of the inputs, respectively. $\Delta \mathbf{u_{\text{min}} }$ and $\Delta \mathbf{u_{\text{max}} }$ are the minimum and the maximum bounds of the inputs rate of change. The matrices $Q$ and $R$ are the weight matrices for states and inputs, respectively.

Within the sampling time $t_s$, the MPC controller should solve the optimization problem in~\eqref{eqn:MPCMathmaticalForm} to find the optimal control inputs and apply the first control input to the system, at every control cycle. However, the solution time of~\eqref{eqn:MPCMathmaticalForm} may exceed $t_s$, especially for long horizon $\mathcal{T}$, and thus the controller gives a lagging input to the system in this case. In the following, we discuss the idea of parametrization that reduces the computational time of solving~\eqref{eqn:MPCMathmaticalForm}.

\subsection*{Parametrization}
\label{sec:Parametrization}
\begin{figure}[t]
\centerline{\includegraphics[width=.45\textwidth]{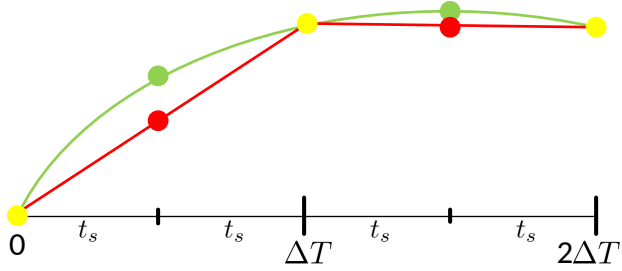}}
\caption{ An example of $\mathcal{T} = 5$ control inputs parameterized by $\mathcal{P} = 3$. The green circles represent the instances at which the control inputs are computed without parametrization. After parametrization, the computations are conducted only at the yellow circles, while the other red points are linearly interpolated.}
\label{fig:parmInputTrj}
\end{figure}

The optimization function in~\eqref{eqn:MPCMathmaticalForm} is solved to find $\mathcal{T}$ vectors of system inputs. By parametrizing the input trajectories with some basis functions, the optimization~\eqref{eqn:MPCMathmaticalForm} is solved but for a fewer number of control inputs. Specifically, the system inputs are estimated only at specific sample instances over the horizon in every control cycle, while the rest is evaluated from the parameterized version. Without loss of generality, we build our discussion here upon the parametrization technique in~\cite{hyatt2020parameterized}, which simply linearizes the trajectories with line segments, as shown in Fig.~\ref{fig:parmInputTrj}.

Let $\mathcal{P}$ represents the number of sample instances over the horizon $\mathcal{T}$ at which the control inputs are estimated from the optimization~\eqref{eqn:MPCMathmaticalForm} (such as the yellow points in Fig.~\ref{fig:parmInputTrj}). The distance between these sample instances in the unit of time steps can be expressed as
\begin{equation*}
\Delta T=\frac{\mathcal{T}-1}{\mathcal{P}-1}.
\label{eq:Delta T}
\end{equation*}
The input $\mathbf{u}_{k}$ can be derived by linear interpolation of the inputs at the next sample $k_n$ and the previous one $k_p$ as
\begin{equation}
\mathbf{u}_{k} = (1-w) \mathbf{u}_{k_p}+(w) \mathbf{u}_{k_n},
\label{parm-u}
\end{equation}
where
\begin{equation*}
\begin{aligned}
k_p &=\left\lfloor\frac{k}{\Delta T}\right\rfloor,\;\;\; k_n = k_p+1, \;\; \textnormal{and } w =\frac{k}{\Delta T} -k_p.
\end{aligned}
\end{equation*}
By modifying the dynamic equation \eqref{eqn:states_vector} by $\mathbf{u}_{k}$ in \eqref{parm-u}, we get 
\begin{equation*}
\mathbf{x}_{k+1}=\mathbf{A}_{d} \mathbf{x}_{k} + (1-w) \mathbf{B}_{d} \mathbf{u}_{k_p}+(w) \mathbf{B}_{d} \mathbf{u}_{k_n},
\end{equation*}
thus, the cost function \eqref{eqn:MPCCost} is modified to be
\begin{equation}
\begin{aligned}
&J'(\mathbf{z}_c, \mathcal{T}, \mathcal{P})=\sum_{k \in \Upsilon}
\left(\left[\refx_k-\mathbf{x}_{k}\right]^{T} \mathbf{Q}\left[\refx_k-\mathbf{x}_{k}\right]\right. \left.+\left[\refu_k-\mathbf{u }_{k}\right]^{T} \mathbf{R}\left[\refu_k-\mathbf{u }_{k}\right]\right),
\end{aligned}
\label{eqn:MPCCostParm}
\end{equation}
s.t. the same constraints as in~\eqref{eqn:MPCCost}, where $\Upsilon = \{ c, c+\Delta T, \dots, c + (\mathcal{P}-1) \Delta T \}$. Now, 
\begin{equation}
    \mathbf{z}_c^{*} = \arg \min_{\mathbf{z}_c} J'(\mathbf{z}_c, \mathcal{T}, \mathcal{P}).
    \label{eqn:MPCMathmaticalForm2}
\end{equation}
This makes the optimization to be solved on $\mathcal{P} < \mathcal{T} $ steps. Moreover, the dimensions of the matrices of the optimization problem are reduced. Specifically, the dimension of $\mathbf{z}_c$ will be ($n\mathcal{P} \times 1$) instead of ($n\mathcal{T} \times 1$) and the matrix $\mathbf{S}_c$ will have a dimension of ($m\mathcal{T} \times n\mathcal{P}$) instead of ($m\mathcal{T} \times n\mathcal{T}$). Therefore, the parametrization reduces the computational time required in each control cycle. However, this reduction in computational time may be at the expense of the MPC performance, especially for small $\mathcal{P}$. This is because the control inputs that are evaluated using~\eqref{parm-u} may be different than the inputs estimated from the original optimization in~\eqref{eqn:MPCMathmaticalForm}. This difference vanishes when $\mathcal{P}$ is close to $\mathcal{T}$.

\section{Motivation and problem statement} 
\label{sec:Motivationandproblemstatement}

As shown in~\eqref{eqn:MPCMathmaticalForm2}, there are two factors that strongly affect both the computational burden and the performance of the MPC, which are the horizon length $\mathcal{T}$ and the sample count $\mathcal{P}$. To enhance the performance of the MPC, both $\mathcal{T}$ and $\mathcal{P}$ are \textbf{fixed} to specific large values for all control cycles. When the trajectories of states have rapid changes over future time, the MPC takes the advantage of large $\mathcal{T}$ and $\mathcal{P}$ for better preparing the appropriate inputs $\mathbf{z}^{*}_c$. This allows the MPC to avoid the overshoot that may occur at these rapid changes, which results in better performance but at the expense of more computational cost. However, fixing $\mathcal{T}$ and $\mathcal{P}$ are not optimal for all trajectories in every control cycle. For example, if the trajectories do not show too many variations, then we can use smaller values for $\mathcal{T}$ and $\mathcal{P}$ to solve \eqref{eqn:MPCMathmaticalForm2} without noticeable loss in control performance but at much less computational cost. This means that it is better to set a variable value for $\mathcal{T}$ and $\mathcal{P}$ for every control cycle according to the variations exhibited in the trajectories. In the rest of the paper, we represent the \textit{variable} horizon length and sample count as $T_c$ and $P_c$, respectively, for the $c^{\text{th}}$ control cycle. 

In the following, we first present an experimental validation that illustrates the effect of different fixed settings for $\mathcal{T}$ and $\mathcal{P}$ on both the MPC's computational time and performance followed by our problem statement that mathematically formulates the problem of adaptive selection of $T_c$ and $P_c$ in every control cycle.

\subsection{Motivating examples}
\label{sec:MotivatingExamples}
We validate our claim on two different models. The first model is a linear vehicle model that is built based on a simple bicycle model approximation of a vehicle and we control the vehicle's lateral position and orientation. The second one is a more complex non-linear robot model that is linearized around the operating point using Linear parameter varying (LPV)\cite{misin2020model} which encapsulates the nonlinearity of the robot model into a linear form. We control the position and orientation of the robot. Both vehicle and robot models are presented with more details in Sec. S.I and Sec. S.II, respectively in the supplementary material. Figure~\ref{fig:motivationRef} shows the reference state trajectories for both the vehicle and the robot. As shown in the figure, two different sets of reference state trajectories are examined, the first set has slow changes (shown in black color) while the second one has rapid changes (shown in green color). 

\begin{figure*}
     	\centerline{\includegraphics[width=.55\textwidth]{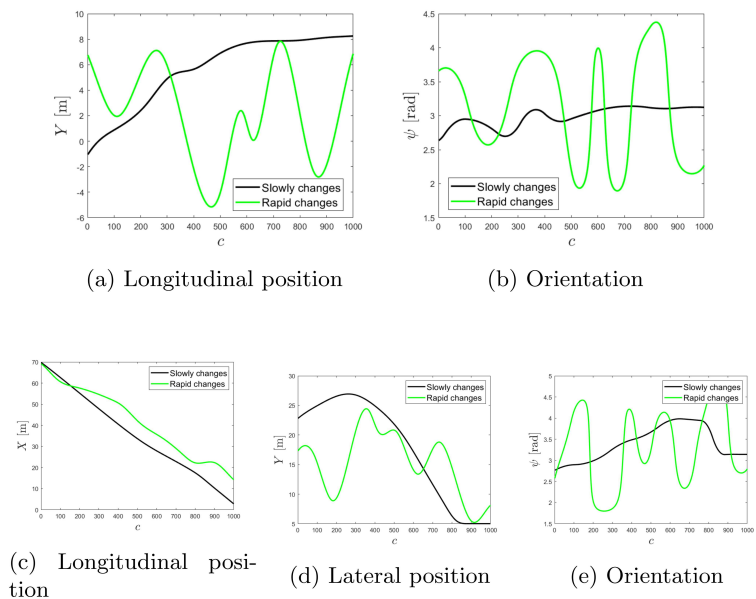}}
        \caption{The vehicle's state reference trajectories are presented in (a) and (b). The robot's state reference trajectories are presented in (c), (d), and (e). The black line represents a reference trajectory that is characterized by slow changes, while the green line represents a reference trajectory that is characterized by rapid changes.}
         \label{fig:motivationRef}
\end{figure*}

We run our experiments using four different settings for $\mathcal{T}$ and $\mathcal{P}$ of the MPC. To differentiate between these settings, we denote MPC($a$,$b$) for a specific setting of $\mathcal{T}$ to $a$ and $\mathcal{P}$ to $b$. The four settings used in our experiments are MPC(40,40), MPC(5,5), MPC(40,3), and MPC(40,25). Thus, the first two settings are concerned with both long and short values for $\mathcal{T}$, respectively, without parametrization, i.e., $\mathcal{P} = \mathcal{T}$. The other two settings are concerned with a long $\mathcal{T}$, but with different settings for $\mathcal{P}$. We evaluate the performance of the MPC with every setting with both the average computational time and the average cost over all control cycles. Assume that we have $H$ control cycles\footnote{We used the same $H$ for all experiments.}, the average cost $E$ is computed as
\begin{equation}
    E = \frac{1}{H} \sum_{c=1}^{H} \frac{1}{\mathcal{T}} J'(\mathbf{z}_c^*,\mathcal{T},\mathcal{P}).
    \label{eq:avgCost}
\end{equation}

\begin{figure*}[htbp]
\centering
\centerline{\includegraphics[width=.75\textwidth]{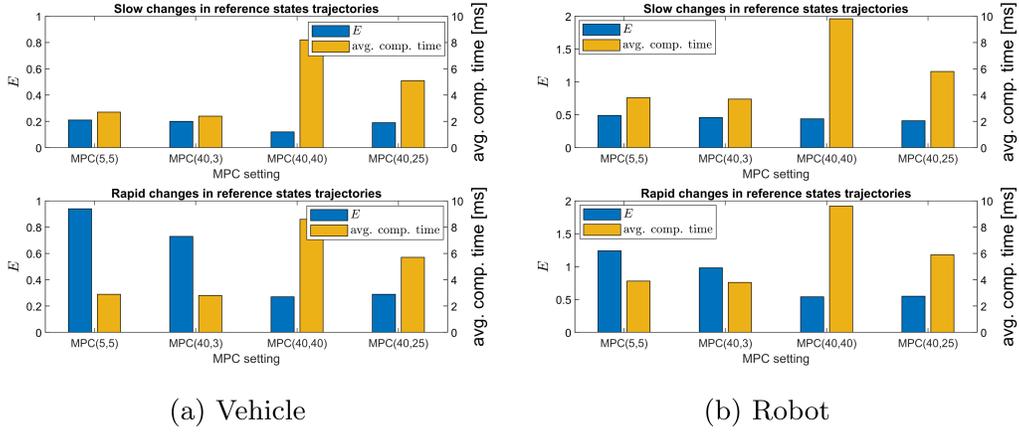}}
\caption{Plot of the average computational time and the average cost against different settings of MPC for (a) the vehicle and (b) the robot. If the reference state trajectories exhibit rapid changes, the average cost is large when using a short horizon MPC(5,5) or using a long horizon with a small sample count of MPC(40,3). For slow variations, the average cost becomes small for all settings. However, the computational time is much smaller for short horizons MPC(5,5) or for a small sample count MPC(40,3).}
\label{fig:motivationResults}
\end{figure*}

In Fig.~\ref{fig:motivationResults}, we plot both the average cost $E$ and the average computational time for the four different settings. As shown in the figure, if the reference state trajectories exhibit rapid changes, the average cost becomes very large when using a short horizon or using a long horizon with a small number of sample counts, but the computational time, in this case, is relatively small. For the same trajectories, when using a long horizon or using a long horizon with a sufficient number of sample counts, the average cost becomes smaller but at the expense of more computational time. In the case of reference state trajectories with slow changes, the average cost becomes small, i.e., good performance, for all different settings used in the MPC even with short horizons or a small number of samples. However, this good performance is obtained in a much smaller computational time when using short horizons or a small sample count.

From the results, we can conclude that despite increasing the horizon length and sample count improving the control performance in general, it is not required in all situations. In real scenarios, the state trajectories may exhibit both rapid and slow variations. Therefore, the horizon length and the sample count should not be fixed and should be adaptively selected according to the variations encountered in the state trajectories in every control cycle. With this adaptive selection, the computational cost of the MPC is reduced, while it maintains its performance unaffected.

\subsection{Problem statement}

\begin{figure*}[htbp]
\centerline{\includegraphics[width=.5\textwidth]{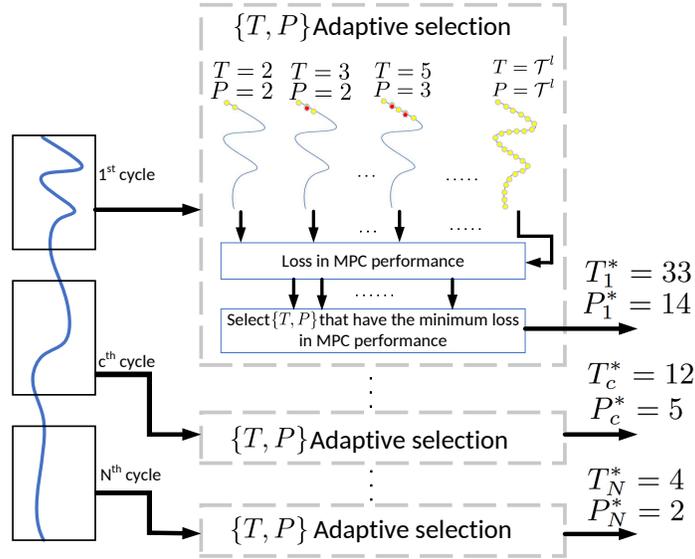}}
\caption{Problem statement architecture: In each $c^{th}$ control cycle, we need to determine what is the most suitable $\{T, P\}$ to be used in MPC optimization.}
\label{fig:problem}
\end{figure*}

A pictorial representation of our problem is shown in Fig.~\ref{fig:problem}. According to the situation of the trajectories in the $c^{\text{th}}$ control cycle, the values of $T_c$ and $P_c$ should be selected as minimum as possible such that the performance of the MPC is not deteriorated compared with its performance when using large values for horizon length and sample count. Let $\mathcal{T}^l$ represents this large value, then our problem can be formulated as 
\begin{equation}
    \begin{aligned}
        \{T_c^{*}, P_c^{*}\} &= \min_{T, P}  \{T, P\}  \\ 
        \text{s.t.}  \\ 
    &  \ell \left(\frac{1}{T} J'(\mathbf{z}_c^{*}, T, P), \frac{1}{\mathcal{T}^l} J'(\gammaB^{*}, \mathcal{T}^l, \mathcal{T}^l) \right) < \epsilon ,   \\ 
    & T \in \{2,\dots,\mathcal{T}^l\},  \; P< T, 
\end{aligned}
    \label{eq:proposed_optimization}
\end{equation}
where $\epsilon$ is a small positive number and $\gammaB^{*}$ is the vector of optimal control inputs that is estimated when using $\mathcal{T}^l$ for both the horizon length and the sample count. In our formulation, we assess the loss in performance of MPC by $\ell$ which measures the relative difference of the average of the cost function $J'$~\eqref{eqn:MPCCostParm} when using the values of $\{T, P\}$ instead of $\mathcal{T}^l$ for both horizon length and the sample count, i.e.,
\begin{equation}
    \ell\left(a, b \right) = \frac{\lvert a-b \rvert}{b}.
    \label{eq:loss}
\end{equation}

A possible solution for~\eqref{eq:proposed_optimization} is that, in every control cycle, we can start with small values of $T$ and $P$ then iteratively increment both values until $\{T_c^{*}, P_c^{*}\}$ are estimated. However, this solution is not practical in run time (i.e., when the controller is in action) as it involves solving the MPC optimization~\eqref{eqn:MPCMathmaticalForm2} several times in every control cycle to obtain $\{T_c^{*}, P_c^{*}\}$. In the following section, we present our proposed efficient approach to estimate $\{T_c^{*}, P_c^{*}\}$.

\section{Proposed adaptive regression-based MPC}
\label{sec:ProposedadaptiveregressionbasedMPC}
\begin{figure}[hbp]
\centerline{\includegraphics[width=.65\textwidth]{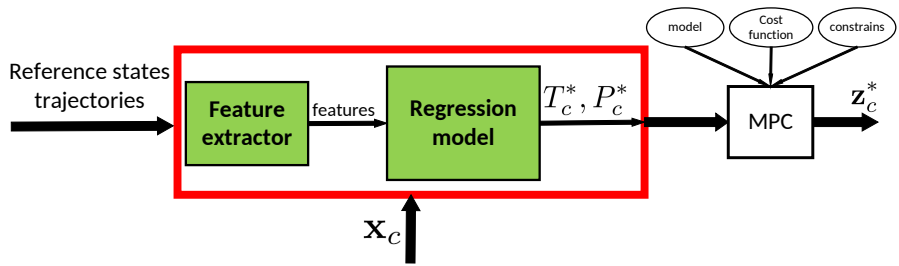}}
\caption{The overall architecture of the proposed regression-based MPC. At each control cycle, reference state trajectories along with the current states of the system being controlled are passed through a feature extractor, then the extracted features are fed to the regression model to predict $T_c^*$ and $P_c^*$ that will be used by the MPC in this control cycle.}
\label{fig:proposedMPC}
\end{figure}

In this section, we present our adaptive regression-based MPC (ARMPC) scheme which adaptively estimate $\{T_c^{*}, P_c^{*}\}$ in every control cycle  from~\eqref{eq:proposed_optimization} using a regression model. Our proposed approach is illustrated in Fig.~\ref{fig:proposedMPC}. Specifically, we build a dataset to train a regression model, then we used the regression model to predict $\{T_c^{*}, P_c^{*}\}$. The predicted $\{T_c^{*}, P_c^{*}\}$ are used to estimate the required control inputs $\mathbf{z}^{*}_c$ from~\eqref{eqn:MPCMathmaticalForm2} as usual. In the following, we discuss the feature extraction, the dataset creation, and the regression model in more detail.

\subsection{Feature Extraction} 
\label{ssec:FeatureExtraction}
Let $\alphaB_c^i = [\refxi^i_{c}, \dots, \refxi^i_{c+\mathcal{T}^l-1}]$ denotes the $i^\text{th}$ reference state trajectory in the $c^\text{th}$ control cycle. We first extract a feature vector $\featureCI$ from each $\segCI$. We designed $\featureCI$ to capture the variation in the reference state trajectory as it is the key element that determines the best minimum values of horizon length and sample count, as we indicated earlier. Specifically, the vector $\featureCI$ composed of
\begin{itemize}
\item \textbf{Curvature} $\mathcal{C}(\segCI)$ is a value that quantifies the amount by which $\segCI$ deviates from being a straight line. Mathematically~\cite{najman2017curvatureIntro},   
\begin{equation*}
\mathcal{C}(\segCI) = [\mathcal{K}(r^1_c), \dots, \mathcal{K}(r^m_c)],
\label{eq:curvatureTotal}
\end{equation*}
where
\begin{equation}
\mathcal{K}(r^i_c) =   \frac{1}{\mathcal{T}^l} \sum_{k=c}^{c+\mathcal{T}^l-1} \frac{\lvert \nabla^2  r_k^i \rvert}{\left(1+(\nabla r_k^i)^{2}\right)^{3 / 2}},
\label{eq:curvature}
\end{equation}
and $\nabla r_k^i = r_{k+1}^i - r_k^i$. It is clear from~\eqref{eq:curvature}, that the curvature value is large for curved trajectories and this value decreases whenever the trajectory shape gets closer to a straight line.

\begin{figure}[t]
\centerline{\includegraphics[width=.55\textwidth]{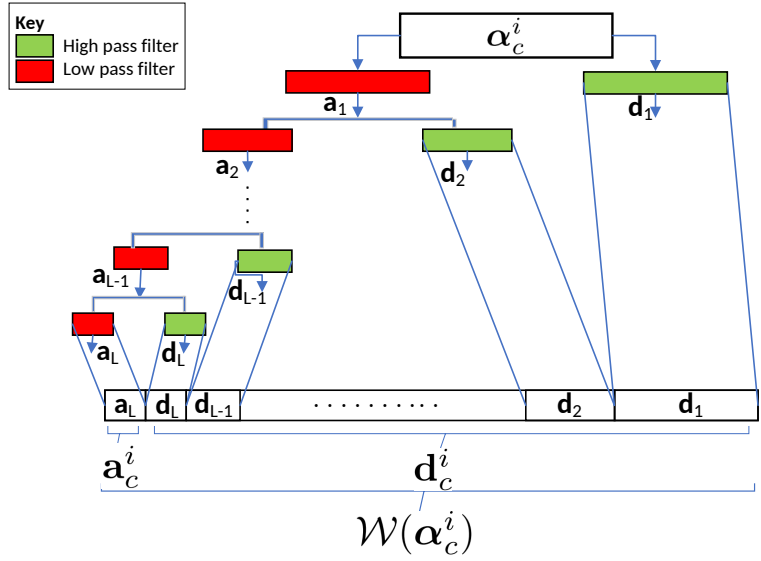}}
\caption{Wavelet decomposition.}
\label{fig:wavletDecomposition}
\end{figure}

\item \textbf{Wavelet coefficients} $\mathcal{W}(\segCI)$ which are extracted from $\segCI$ using the wavelet decomposition described in the pyramid architecture in Fig.~\ref{fig:wavletDecomposition}. Specifically, two sets of coefficients are computed from $\segCI$: approximation coefficients $\mathbf{a}^{c}_{i}$ and detail coefficients $\mathbf{d}^{c}_{i}$~\cite{wavelet:Mallat:1989}. The approximation and the detail coefficients are computed by convolving $\segCI$ with a low pass filter and a high pass filter, respectively, followed by dyadic decimation. The same procedure is repeated $L$ times with the approximation coefficients and all resultant coefficients are used to form $\mathcal{W}(\segCI)$. The wavelet decomposition is used in our features because it localizes the trajectory in both time and frequency~\cite{wavelet:Mallat:1989}. Thus, a stretched wavelet helps capture the slowly varying changes in the state trajectory while a compressed wavelet helps capture abrupt changes in the state trajectory.

\end{itemize}
Thus, 
\begin{equation}
\featureCI = [\mathcal{C}(\segCI),\mathcal{W}(\segCI)].
\label{eq:feature_vec}
\end{equation}

Besides capturing the variations in the reference state trajectory, we augment our features by the error vector $\mathbf{e}_c = [e^1_c, \dots, e^m_c]$, where $e^i_c = \lvert \refxi^i_{c} - x^i_c \rvert$ represents the absolute error between the values of the reference state and the current state at the time instant $c$. The reason for taking the error into account is that the error has a noticeable effect on the selection of horizon length and sample count. When the error is large, it is preferred to use large values for both horizon length and sample count to perfectly derive the states to their reference.

After we compute $\featureCI$ from each $\segCI$, we concatenate the vectors $\featureCI$ for all $i = 1,\dots,m$ along with the error vector $\mathbf{e}_c$ to form the complete feature vector $\featureC$, i.e.,
\begin{equation}
\featureC = [\betaB^{1}_{c}, \dots, \betaB^{m}_{c}, \mathbf{e}_c ]^{T}.
\label{eq:whole_feature_vec}
\end{equation}

To generate the reference states trajectories for feature extraction, we synthetically utilize the discrete state space equation~\eqref{eqn:statespaceform} for a given model by randomly varying the control inputs, while satisfying their constraints.

\subsection{Dataset Creation}
\label{ssec:dataset}
We build the dataset by associating each feature vector $\featureC$ of the $c^\text{th}$ control cycle with its corresponding values of the best minimum horizon $T_c^{*}$ and sample count $P_c^{*}$. Algorithm~\ref{alg:datasetCreation} illustrates how $\{T_c^{*}, P_c^{*}\}$ are obtained. Specifically, our dataset is built in two successive steps. First, we estimate $T_c^{*}$, then we estimate $P_c^{*}$ for the specific estimated $T_c^{*}$. To estimate $T_c^{*}$, we initially set both the horizon length and the sample count to a small value $T$, then iteratively increment $T$ and solve~\eqref{eqn:MPCMathmaticalForm2} with both horizon length and sample count set to $T$. In each iteration, the loss in performance~\eqref{eq:loss} of the MPC is computed. If the loss falls below a threshold $\epsilon$, $T_c^{*}$ is estimated and the iterations are stopped. Otherwise, $T_c^{*}$ is set to the maximum allowable horizon length $\mathcal{T}^{l}$. Once we obtain $T_c^{*}$, we estimate $P_c^{*}$ in the next step. We set the sample count to a small value $P$, then iteratively increment $P$ and solve~\eqref{eqn:MPCMathmaticalForm2} with horizon length equal to $T_c^{*}$ and sample count set to $P$. In each iteration, the loss~\eqref{eq:loss} is computed and if it falls below $\epsilon$, the value of $P_c^{*}$ is returned.

\begin{algorithm}
\caption{Estimation of $T_c^*$ and $P_c^*$ for the $c^{\text{th}}$ control cycle for dataset creation.}\label{alg:datasetCreation}

    \hspace*{\algorithmicindent} \textbf{Input: {$\mathcal{T}^l, \epsilon$}} \\
    \hspace*{\algorithmicindent} \textbf{Output: {$T_c^*,P_c^*$}} 
\begin{multicols}{2}
\begin{algorithmic}[1]
  \State $T \gets 2$
  \State $a \gets  \frac{1}{\mathcal{T}^l} J'(\gammaB_c^*, \mathcal{T}^l, \mathcal{T}^l)$
  \While{$T<\mathcal{T}^l$}
     \State $b \gets  \frac{1}{T} J'(\mathbf{z}_c^*, T, T)$   
     \If{$ \ell(b,a) < \epsilon $}
        \State $T_c^{*} \gets T$
        \State break
     \EndIf
    \State $T \gets T+1$
  \EndWhile
  \State $T_c^{*} \gets T$
  \State $P \gets 2$
  \While{$P<T_c^*$}
  \State $c \gets \frac{1}{T_c^*} J'(\mathbf{z}_c^*, T_c^*, P)$    \If{    $ \ell(c,a)  < \epsilon$}
        \State $P_c^{*} \gets P$
       \State break
     \EndIf
  \State $P \gets P+1$
  \EndWhile
  \State $P_c^{*} \gets P$
\end{algorithmic}
\end{multicols}
\end{algorithm}

The Algorithm~\ref{alg:datasetCreation} has high computational cost as it involves solving the MPC optimization~\eqref{eqn:MPCMathmaticalForm2} several times to obtain $\{T_c^*, P_c^*\}$. However, all these computations are performed offline, i.e., in the training time. But, once the dataset is built, we predict the values of $\{T_c^*, P_c^*\}$ in run time using very simple calculations, thanks to the regression model that we discuss next. 

\vspace*{-0.05in}
\subsection{Regression Analysis}
\label{ssec:RegressionAnalysis}
\begin{figure}
\centerline{\includegraphics[width=.65\textwidth]{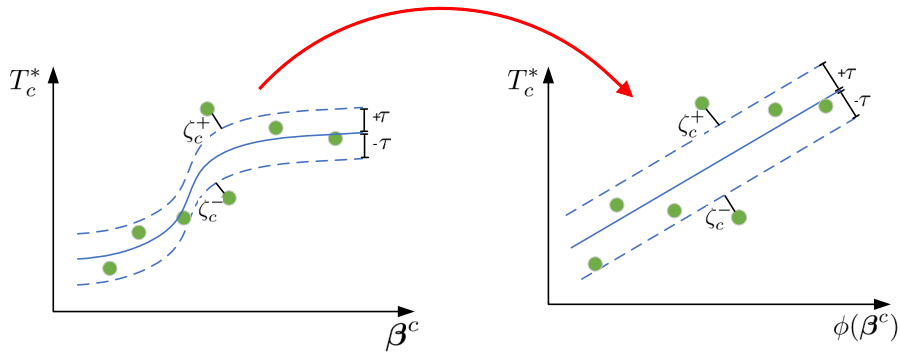}}
\caption{The kernel function $\phi$ transforms the data into a higher dimensional feature space to make it possible to find a linear hyperplane that holds maximum number of training observation within the margin.}
\label{fig:SVR}
\end{figure}

Our training dataset contains $N$ records, one for each control cycle. Every record is composed of the feature vector $\featureC$ for the $c^\text{th}$ control cycle with its corresponding values of $\{T_c^{*}, P_c^{*}\}$. Our goal here is to build two regression models, the first is for predicting the horizon length and the second one is for predicting the sample count, from the given feature vector. We used the non-linear support vector regression (SVR)~\cite{gunn1998SVM} technique for building our regression models because of its robustness to outliers, excellent generalization capability, and high prediction accuracy. The objective of the SVR is to find the hyperplane $\omegaB^{T} \phi(\featureC) + b$ that holds maximum training observation within the margin $\tau$ (tolerance level), as shown in Fig.~\ref{fig:SVR}, where $\phi(\featureC)$ is a transformation that maps $\featureC$ to a higher-dimensional space. Mathematically, the goal of the training of the SVR model is to find the best parameters $\omegaB^{*}$ for the hyperplane by solving 
\begin{equation}
\begin{aligned}
  \{\omegaB^{*}, b^{*} \} =  &\arg \min_{\omegaB, b}   \frac{1}{2} \| \omegaB \| ^{2} + C \sum_{c=1}^{N} ( \zeta^{+}_c + \zeta^{-}_c ), \\
   &\text{s.t.}\\
   & \forall c:  T_c^{*} - \left(\omegaB^{T} \phi(\featureC)+b\right) \leq \tau +  \zeta^{+}_c ,\\
   & \forall c:  \omegaB^{T}\phi(\featureC) + b - T_c^{*} \leq \tau +  \zeta^{-}_c ,\\
   & \forall c: \zeta^{+}_c, \zeta^{-}_c \geq 0,
\end{aligned}
\label{eq:svr}
\end{equation}
where $C > 0$ is a regularization parameter that penalizes the number of deviations larger than $\tau$. The $\zeta^{+}_c$ and $\zeta^{-}_c$ are slack variables that allow the regression to have errors, as shown in Fig.~\ref{fig:SVR}. 

We used the same formulation~\eqref{eq:svr} to train the second SVR model that predicts the sample counts but with a slight modification for the constraints. Since we estimate $P_c^*$ in our dataset creation for a specific $T_c^{*}$, as shown in Sec.~\ref{ssec:dataset}, we augment the feature vector $\featureC$ with $T_c^{*}$ and treat $P_c^{*}$ as the required output.

The training of the regression in~\eqref{eq:svr} is solved using the dual problem form as shown in~\cite{vapnik1999dualformula} with a Gaussian kernel $K(\mathbf{x}, \mathbf{y}) = \exp\left( -\|\mathbf{x} - \mathbf{y} \|^2 \right) = \phi(\mathbf{x})^{T}\phi(\mathbf{y}) $. We used the cross-validation method~\cite{chorowski2014crossvalidation} with grid search to estimate the values of the hyper-parameters $C$, $\tau$, $\zeta^{+}_c$, and $\zeta^{-}_c$. As we mentioned, we build two regression models, the first is for predicting the best horizon length $T_c^*$, and the second one is for predicting the best sample count $P_c^*$. The output from the training is the best parameters $\{\omegaB^*_{t}, b^{*}_{t} \}$ and $\{\omegaB^*_{p}, b^{*}_{p} \}$ of the hyperplanes of the SVRs corresponding to both $T_c^*$ and $P_c^*$, respectively. The complete control process of the proposed ARMPC is outlined in Algorithm~\ref{alg:wholeprocess}.

\begin{algorithm}
\caption{Adaptive Regression-based Model Predictive Control}
\begin{algorithmic}[1]
	\State Create the dataset as in Algorithm~\ref{alg:datasetCreation}.
	\State Train two SVRs by solving Eq.~\eqref{eq:svr} to estimate $\{\omegaB^*_{t}, b^{*}_{t} \}$ and $\{\omegaB^*_{p}, b^{*}_{p} \}$.
	\State $c$ $\gets$ 0.
	\While{$c< H$}
      \State Compute the feature vector $\featureCI$ for each $\alphaB_c^i$ using Eq.\eqref{eq:feature_vec}.
      \State Compute the error vector $\mathbf{e}_c \gets [e^1_c, \dots, e^m_c]$, where $e^i_c = \lvert \refxi^i_{c} - x^i_c \rvert$.
      \State Compute the feature vector $\featureC$ using Eq.\eqref{eq:whole_feature_vec}.
      \State $T_c^*$ $\gets$ $\omegaB^{*T}_{t}\phi(\featureC) + b^{*}_{t}$.
      \State $P_c^*$ $\gets$ $\omegaB^{*T}_{p}\phi([\featureC, T_c^*]) + b^{*}_{p}$.
	  \State Solve $\mathbf{z}_c^* \gets \arg \min_{\mathbf{z}_c}  J'(\mathbf{z}_c, T_c^*, P_c^*)$.
	  \State Get the first $n$ elements from $\mathbf{z}_c^*$ as the optimal input $\mathbf{u}_c \gets \mathbf{z}_c^* [1:n]$.
	  \State Apply the optimal input $\mathbf{u}_c$ and update states $\mathbf{x}_{c+1} \gets \mathbf{A}_d \mathbf{x}_{c}+ \mathbf{B}_d \mathbf{u}_{c}$.
	  \State $c \gets c+1$.
  	\EndWhile
     
\end{algorithmic}
\label{alg:wholeprocess}
\end{algorithm}

\section{Experimental results}
\label{sec:Experimentalresults}
We evaluate our proposed ARMPC on linear vehicle and non-linear robot models used in Sec.~\ref{sec:MotivatingExamples}. We are interested in controlling both position and orientation for both models. For the vehicle, we control the lateral position only, while for the robot, we control both longitudinal and lateral positions, assuming the states of both models are measurable to the MPC.

Independent of our experiments, we build the dataset to train our regression model as we discussed in Sec.~\ref{ssec:dataset}. We set the parameters of Algorithm~\ref{alg:datasetCreation} as $\epsilon = 0.05$ and $T^{l} = 40$. The generated training datasets contain $N = 2000000$ for both models. Also, we used the Daubechies wavelet with order 2~\cite{daubechies1992tenLecWaveletBook} to perform the wavelet decomposition with $L=3$ levels. The parameters of our regression model are estimated with 5-fold cross-validation and their values are $C = 7.4129$, $\tau=0.62$, and $\zeta^{+}_c = \zeta^{-}_c = 0.1$. Figure~S.3 in the supplementary material shows the reference spatial trajectories for both the vehicle and the robot. Also, Fig.~S.4 in the supplementary material shows their reference state trajectories, which as in the reality, they may contain both rapid and slow variations as shown in the figure. In all conducted experiments, the output measurements from both models are obtained with added noise to simulate the modeling errors and/or disturbances that may occur in a real scenario. An extended Kalman filter~\cite{Bang2018kalman} is used in state estimation. All experiments are carried out for 40 seconds. We used Matlab/Simulink software running under windows 10, with a PC (i7-8550U CPU @1.80 GHz, 16 GB RAM).

In the following three subsections, we first present the performed experiments that compare the proposed ARMPC with other state-of-the-art techniques. Then, we present an ablation study that compares the proposed ARMPC with defeatured versions obtained by introducing some modifications to the proposed technique. Finlay, we discuss the behavior of the proposed ARMPC when encountering a non-optimal control situation that may arise due to the in-feasibility of solving the MPC optimization before the control cycle time is over or due to system characteristics and constraints.

\subsection{Comparison with other techniques}
\label{ssec:comparOthertechs}

We compare the proposed ARMPC with three different state-of-the-art MPC techniques, which are the parametrized MPC (PMPC)~\cite{hyatt2020parameterized}, the adaptive dual MPC (ADMPC)~\cite{kim2021dual}, and the adaptive neural network MPC (ANMPC)~\cite{gardezi2018neuralNetwork} techniques. Additionally, we include the standard MPC (SMPC) with a long horizon as a baseline for performance. We used the same notations in Sec.~\ref{sec:MotivatingExamples} to denote both the PMBC and SMPC, which are MPC(40,$\mathcal{P}$) and MPC(40,40), respectively. Our comparison is performed using two experiments. In the first experiment, we tune the parameters of every technique to provide the best performance so we can compare the amount of speed-up of each technique with the baseline. In the second experiment, we tune the parameters to give the same speed-up as our proposed ARMPC so we can compare the performance of each technique. The parameters of all techniques for both experiments are listed in Sec. S.III in the supplementary material. To ensure a fair comparison, the parameters of the MPC's optimization~\eqref{eqn:MPCCostParm} and constraints settings in all compared techniques are kept the same as shown in Table S.II and Table S.III, respectively, in the supplementary material. We used the quadprog~\cite{geletu2007quadprog} convex solver for solving the MPC's optimization~\eqref{eqn:MPCCostParm} for all techniques. For every technique, we measure the time span from the start of every control cycle till the estimation of the control inputs, then we report the average computational time across all cycles.

\begin{figure}[t]
\centerline{\includegraphics[width=.65\textwidth]{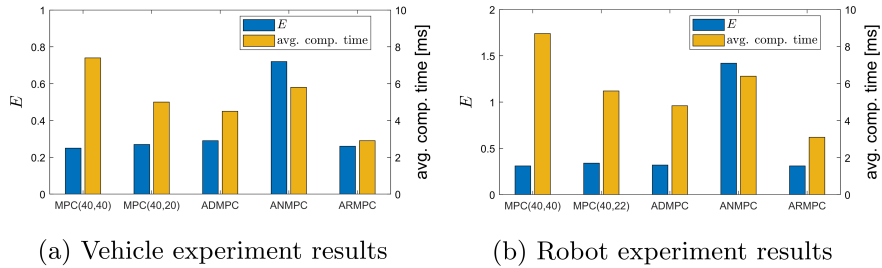}}
\caption{The results of the first experiment in which we tune the parameters of every technique to provide the best performance. The proposed ARMPC provides a much smaller average computational time with comparable performance to the MPC with long horizon MPC(40,40). Also, the proposed ARMPC shows a 35\%-52\% reduction in the average computational time compared with the other techniques without loss in performance.}
\label{fig:expResultsPerformance}
\end{figure}

The results\footnote{A video that shows simulation for the vehicle controlled by the proposed ARMPC is shown in Sec. S.IV in the supplementary material.} of the first experiment are shown in Fig.~\ref{fig:expResultsPerformance}. Specifically, we plot both the average cost $E$ in ~\eqref{eq:avgCost} and the average computational time for all techniques under comparison. As shown in the figure, the proposed ARMPC provides a much smaller average computational time with comparable performance to the MPC(40,40). Specifically, the proposed ARMPC reduces the average computational time by 61\% and 64.5\% compared with MPC(40,40), for the vehicle and the robot, respectively. Also, the proposed ARMPC shows a 35\%-52\% reduction in the average computational time compared with the other techniques without loss in performance. This reduction in computational time is obtained because our proposed ARMPC does not estimate the horizon length only and it does not rely on raw values of the reference states but employs feature engineering to prevent overfitting in the learning phase. Additionally, the proposed ARMPC extracts these features over a future span of the horizon which leads to a more accurate estimation. Conversely, both ADMPC and PMPC techniques fix the sample count so it is required to use a large enough sample count to give good performance in all situations which reflects in the overall average computational time. Also, the ANMPC technique fixes the ratio between the predicted sample count and the time span of the horizon, uses only the instantaneous values of the states, and does not employ any feature engineering for making predictions. These drawbacks make the technique provide a poor estimation of the sample count which is reflected in the figure where the ANMPC technique shows the poorest performance with high computational time.

\begin{figure*}[t]
\centerline{\includegraphics[width=.55\textwidth]{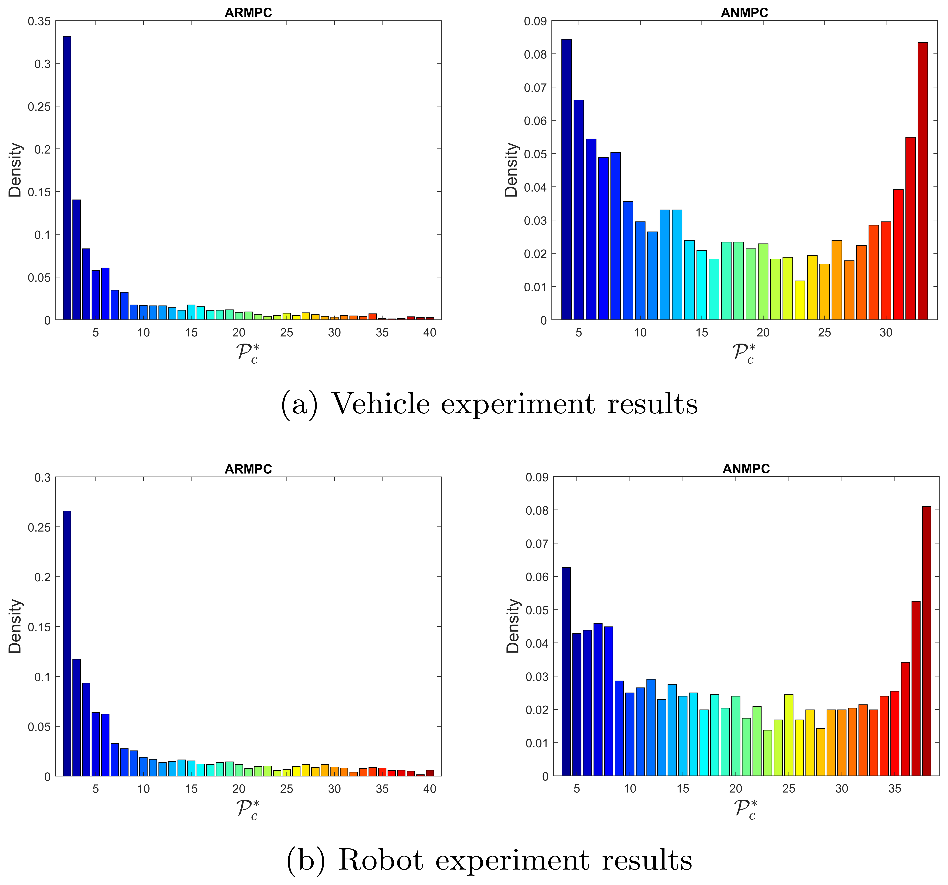}}
\caption{Density distribution for the estimated sample count by the ARMPC and ANMPC techniques. The proposed ARMPC provides a smaller number of occurrences of a large sample count, while the ANMPC technique provides large number of occurrences of a large sample count due to the wrong estimation.}
\label{fig:EXP1_Hist_P}
\end{figure*}

In Fig.~\ref{fig:EXP1_Hist_P}, we plot the histograms of the predicted sample count that is obtained by the proposed ARMPC and the ANMPC techniques for the first experiment. The histograms show that the proposed ARMPC provides a smaller number of occurrences of a large sample count which reinforces the conclusion drawn from Fig.~\ref{fig:expResultsPerformance} that the proposed ARMPC has a smaller average computational time compared with the ANMPC technique. Note that, we omit the other techniques from the histograms because these techniques do not estimate variable sample count.

\begin{figure}[t]
\centerline{\includegraphics[width=.65\textwidth]{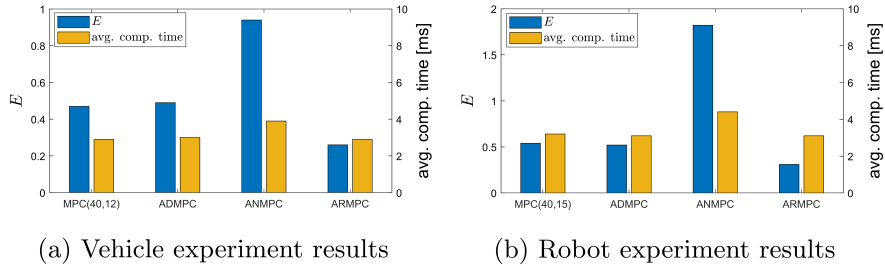}}
\caption{The results of the second experiment in which we tune the parameters of each technique to give the same speed-up as our proposed ARMPC. All techniques fail to provide comparable performance with the proposed ARMPC for the same average computational time.}
\label{fig:exp222Results}
\end{figure}

The results of the second experiment are shown in Fig.~\ref{fig:exp222Results} where we plot both $E$ and the average computational time for all techniques. As shown in the figure, all techniques fail to provide comparable performance with the proposed ARMPC for the same average computational time. For the PMPC technique, we decrease its sample count till we reach the same average computational time as our proposed ARMPC. However, this negatively affects the performance, since there are some situations (states with large variations), and the MPC needs enough samples to represent the horizon length. Also, for the ADMPC technique, modifying the number of dense samples, the number of the sparse sample or the heuristic threshold leads to speeding up the computational time but at expense of the performance. The ANMPC technique, as shown in the first experiment, provides a poor estimation of the sample count. So to enforce reducing its computation, we limit the range of the sample counts that are used in its training in this experiment and thus the technique always predicts a small sample count to save computations. However, as shown in the figure, the performance is negatively affected. 

\subsection{Ablation study: comparison with defeatured versions}
In this section, we study the effect of each feature proposed in Sec.\ref{ssec:FeatureExtraction} on the overall performance and computational time of the proposed ARMPC. Specifically, we compare the proposed ARMPC with three de-featured versions. Each de-featured version is obtained by dropping one of the proposed features from the training of the SVR and consequently from the online prediction. We denote (ARMPC$^{-}$W), (ARMPC$^{-}$C), and (ARMPC$^{-}$E) for the de-featured versions of the proposed ARMPC that is obtained by dropping the wavelet, the curvature, and the error features, respectively. Additionally, we study the effect of predicting only the horizon length (without the sample count). We denote (ARMPC$^{-}$P) for the de-featured version that trains only one SVR to predict the horizon length only. For this version, we always set $\mathcal{P}=\mathcal{T}$.

\begin{figure}[htbp]
\centerline{\includegraphics[width=.65\textwidth]{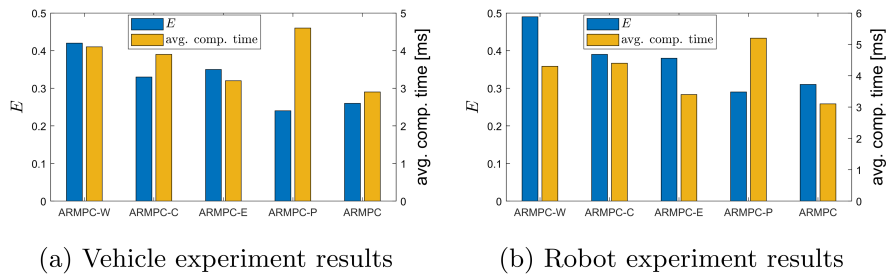}}
\caption{Plot of the average computational time and the average cost of the proposed ARMPC in comparison with four defeatured versions. Drooping any of the proposed features has a negative effect on the performance and increases the computational time due to the wrong estimation of optimal $\{\mathcal{T},\mathcal{P}\}$.}
\label{fig:AblationStudy}
\end{figure}

Figure~\ref{fig:AblationStudy} shows $E$ and the average computational time of the proposed ARMPC in comparison with the four defeatured versions. As shown, predicting the horizon length only without estimating the best minimum sample count as in ARMPC$^{-}$P increases the average computational time with no significant enhancement in the performance. Thus, estimating both the sample count and the horizon length has a very important impact on the computational time reduction. Also, dropping any feature from the proposed features affects the quality of the estimation which is negatively reflected in the performance. This is because the dropped features make the rest indistinctive and in this case, the dataset may have records with the same feature values but with two different associated values of $\{\mathcal{T},\mathcal{P}\}$. For example, suppose that there is a situation where the reference state trajectories have slow variations and small curvature. If the instantaneous error is large in this case, then Algorithm~\ref{alg:datasetCreation} will choose a large value for $\{\mathcal{T},\mathcal{P}\}$ and vice-versa, even with such slow variations and small curvature. Thus if the error feature is dropped, then several records will appear in the dataset with similar values of the curvature and wavelet features but with different associated $\{\mathcal{T},\mathcal{P}\}$. This results in low fitting for the SVR and will lead to non-optimal predictions. Similarly, dropping the wavelet features decreases the performance as the remaining features can not capture whether the state trajectories have rapid or slow variations. Therefore, a dataset record may be constructed with a small value of ${\mathcal{T},\mathcal{P}}$ if the instantaneous error is small regardless of the variations encountered in the trajectories. So we can conclude that all parts of the proposed features are very important for obtaining the best performance in the least amount of computational time. 

\subsection{Infeasibility and non-optimal control}
\label{ssec:nonOptimal}

In this subsection, we discuss the behavior of the proposed ARMPC when encountering a non-optimal control situation that may arise due to the in-feasibility of solving the MPC optimization before the control cycle time is over or due to system characteristics and constraints.

\begin{figure}[htbp]
\centerline{\includegraphics[width=.65\textwidth]{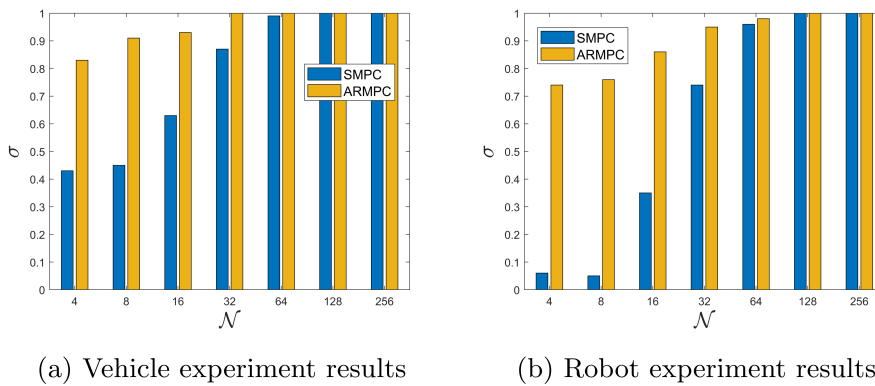}}
\caption{A plot of the percentage of converging to the optimal solution $\sigma$ for different values of the MPC solver's allowable maximum number of iterations $\mathcal{N}$. For large values of $\mathcal{N}$, both proposed ARMPC and SMPC approaches show high values of $\sigma$. However, For small values of $\mathcal{N}$, ARMPC shows a significantly higher $\sigma$ than SMPC. This is because the proposed ARMPC adaptively estimates the best minimum horizon in every control cycle, and therefore, it has smaller calculations involved in finding the optimal solution.}
\label{fig:nonOptimal}
\end{figure}

When the in-feasibility happens because the optimization incurs high computational costs due to a large horizon and can not find the optimal solution before the control cycle time is over, the optimization returns a sub-optimal solution. To compare our proposed ARMPC approach with the SMPC approach, we conduct the following experiment on both the vehicle and the robot models. We limit the maximum number of iterations $\mathcal{N}$ the MPC solver allows to find the optimal solution and quantify the percentage of the number of times both the ARMPC and the SMPC approaches converge to the optimal solution from the total number of control cycles $H$. We denote this percentage by $\sigma$. We repeat the same experiment by varying $\mathcal{N}$ and record the associated $\sigma$ for both ARMPC and SMPC approaches and plot $\sigma$ against $\mathcal{N}$ in Fig.~\ref{fig:nonOptimal}. As shown in the figure, when $\mathcal{N}$ is large, both ARMPC and SMPC converge to the optimal solution often. However, when $\mathcal{N}$ is smaller, the proposed ARMPC shows a significantly higher percentage of converging to the optimal solution. This is because the ARMPC approach estimates the optimal $\{T_c^*, P_c^*\}$, and therefore, it has smaller calculations involved in finding the optimal solution. Thus, the proposed ARMPC approach shows a significant advantage over SMPC in this case.

In the second case, when the in-feasibility happens due to system characteristics and constraints (for example, when the constraints are too tight), we compare the behavior of the ARMPC and the SMPC approaches through the following experiment. We alter the two models' characteristics by adding more tightened constraints to the MPC optimization problem to force the optimization problem to be infeasible. The added constraints are presented for both the vehicle and the robot models in more detail in Table~S.IV in the supplementary material. In figure~\ref{fig:infeasibility}, we plot $\sigma$ for different values of $\mathcal{N}$. The figure shows that both ARMPC and SMPC fail to get acceptable $\sigma$ values. Specifically, neither has a value of $\sigma$ greater than 2\% even with higher values of $\mathcal{N}$. This is simply because the solution can not be found due to the MPC's optimization problem infeasibility.

\begin{figure}[htbp]
\centerline{\includegraphics[width=.65\textwidth]{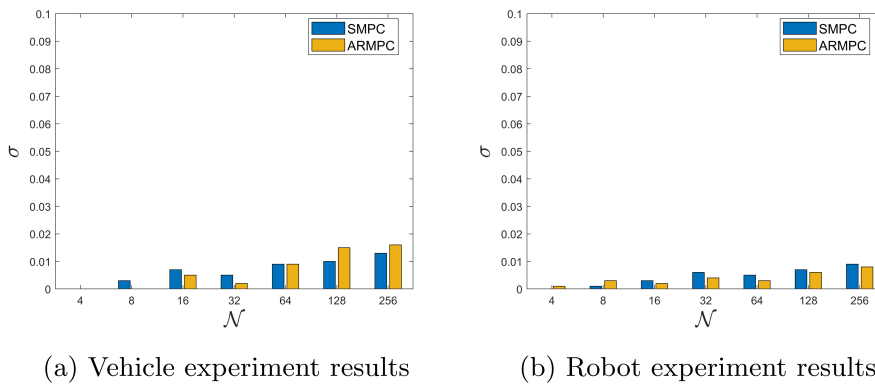}}
\caption{A plot of the percentage of converging to the optimal solutions $\sigma$ for different values of solver's allowable maximum number of iterations $\mathcal{N}$ in case of a tightly constrained optimization problem. Both ARMPC and SMPC fail to show an acceptable percentage of optimal solutions due to the infeasibility of the optimization problem.}
\label{fig:infeasibility}
\end{figure} 

\section{Conclusion and future work}
\label{sec:Conclusion}
In this paper, we propose an adaptive regression-based MPC (ARMPC) technique that predicts the best minimum horizon length and the sample count from several features extracted from the reference state trajectories of the system being controlled. The features are designed to capture the variation in the trajectories by using the wavelet decomposition coefficients and the curvature value, in addition to the instantaneous error between the reference state and the current state. We conducted several experiments on both linear and non-linear models to compare the proposed technique with three different state-of-the-art techniques. The results show that the proposed technique provides a superior reduction in computational time with a reduction of about 35-65\% compared with the other techniques without introducing a noticeable loss in performance. Additionally, we showed experimentally that dropping any of the proposed features makes our regression model not provide an accurate estimation for the best minimum horizon length and the sample count which affects both the performance and the computational time.

In the future, we plan to apply the proposed approach to non-linear MPC with non-linear MPC solvers, such as the genetic algorithm. Another direction is to apply machine learning techniques to the genetic algorithm to adaptively select the best parameters according to some features extracted from the reference and state trajectories.

\vfill


\end{document}


\title{Supplementary Material for ``Fast Adaptive Regression-Based Model Predictive Control''}
\author{Eslam~Mostafa,
        Hussein~A.~Aly,
        Ahmed~Elliethy
}

\maketitle
\IEEEpeerreviewmaketitle

This document provides supplementary material, for the paper~\cite{FastRegMPC:Eslam:2022}. In Sec.~\ref{sec:S.vehicleMdl} and Sec.~\ref{sec:S.robotMdl}, we present the mathematical model of the vehicle and the robot dynamics, respectively, along with the values of the parameters used in our experiments. Section~\ref{sec:S.Expsettings} presents the settings of the adaptive dual MPC (ADMPC) and the adaptive neural network MPC (ANMPC), respectively, with their parameter settings used in the experiments in the main manuscript. Finally, Sec.~\ref{sec:S.video} presents a video showing a simulation of the vehicle controlled by the proposed technique.

\section{Mathematical modeling of vehicle dynamics}
\label{sec:S.vehicleMdl}
\begin{figure}[]
 \centerline{\includegraphics[width=.4\textwidth]{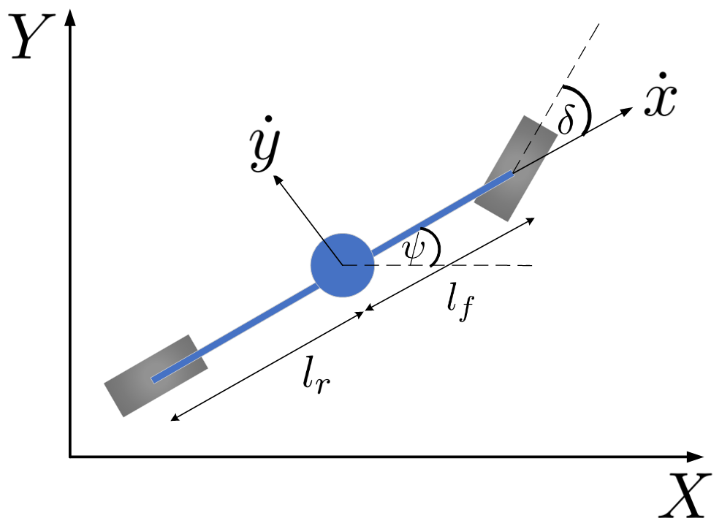}}
 \caption{Bicycle Model}
 \label{fig:bicycleModel}
 \end{figure}

We used the bicycle model~\cite{rajamani2011myvehiclemodel} to represent the vehicle model, as shown in Fig.~\ref{fig:bicycleModel}.
Specifically, two reference frames will be considered in designing the bicycle model: a fixed earth reference frame (E-frame) and a fixed Body frame (B-frame). The E-frame has axis $X$, which points to the North, and axis $Y$ which points towards the East. The B-frame is attached to the body itself. Thus, the B-frame is suitable for velocity measurements, while the E-frame is suitable for position control. According to the equation of motion, the bicycle model is expressed as follows~\cite{rajamani2011myvehiclemodel}

\begin{equation}
\begin{aligned}
&m(\Ddot{y}+\dot{\psi\dot{x}})=F_{yf}+F_{yr} \\
&I_{z}\ddot{\psi}=l_{f} F_{yf}-l_{r} F_{yr},\\
\end{aligned}
\label{eqn:EquationofMotion}
\end{equation}
where $m$ denotes the vehicle mass, $\dot{y}$ is the lateral velocity of the vehicle, $\dot{\psi}$ is the yaw rate, $\dot{x}$ denotes the longitudinal velocity of the vehicle, $I_{z}$ is the moment of inertia about the z-axis,  $l_{f}$ and $l_{r}$ are the distances from the center of gravity of the vehicle to the front and rear axles, respectively. According to bicycle model assumptions, the force applied to each tire is written as follows~\cite{rajamani2011myvehiclemodel}
\begin{equation}
\begin{aligned}
&F_{yf}=2 C_{f} \alpha_{f}=2 C_{f}(\delta-\theta_{vf}) \\
&F_{yr}=2 C_{r} \alpha_{r}=2 C_{r}(-\theta_{vr}),
\end{aligned}
\label{eqn:vehicleforces}
\end{equation}
where $F_{yf}$ and $F_{yr}$ are the lateral tire forces of the front and rear axles, respectively, and $C_f$ and $C_r$ are the cornering stiffness of the front and rear tires, respectively. $\alpha_f$ and $\alpha_r$ are the slip angles of the front and rear tires, respectively, and $\delta$ is the steering angle. $\theta_{vf}$ and $\theta_{vr}$ are the velocity angles of the front and rear tires. Substituting~\eqref{eqn:vehicleforces} into~\eqref{eqn:EquationofMotion}, vehicle lateral dynamics models are expressed as
\begin{equation}
\begin{aligned}
\ddot{y}=&-\left(\frac{2 C_{f}+2 C_{r}}{m V_{x}}\right) \dot{y}-\left(V_{x}+\frac{2 C_{f} l_{f}+2 C_{r} l_{r}}{m V_{x}}\right) \dot{\psi} \\
\ddot{\psi}=&-\left(\frac{2 C_{f} l_{f}-2 C_{r} l_{r}}{I_{z} V_{x}}\right) \dot{y} \\
&-\left(\frac{2 C_{f} l_{f}^{2}+2 C_{r} l_{r}^{2}}{I_{z} V_{x}}\right) \dot{\psi}+\frac{2 C_{f} l_{f}}{I_{z}} \delta.
\end{aligned}
\label{eqn:vehiclemodel}
\end{equation}
Since our reference trajectory lateral movement is given in the earth reference frame, we need to relate the lateral velocity in the vehicle body frame to the earth frame. This can be done by
\begin{equation}
\dot{Y}=\dot{y}\cos{\psi}+\dot{x}\sin{\psi},\\
\label{eqn:extraeqn}
\end{equation}
where $Y$ is the vehicle's lateral position in the Earth reference frame.

From~\eqref{eqn:vehiclemodel} and \eqref{eqn:extraeqn}, the vehicle system can be represented as the linear time-invariant (LTI) system in state-space form as 
\begin{equation}
{\left[\begin{array}{c}
\ddot{y} \\
\dot{\psi} \\
\ddot{\psi} \\
\dot{Y}
\end{array}\right]  =\left[\begin{array}{cccc}
a & 0 & b & 0 \\
0 & 0 & 1 & 0 \\
c & 0 & d & 0 \\
1 & \dot{x} & 0 & 0
\end{array}\right]\left[\begin{array}{c}
\dot{y} \\
\psi \\
\dot{\psi} \\
Y
\end{array}\right]} 
+\left[\begin{array}{c}
e  \\
0 \\
f \\
0 
\end{array}\right]\left[\begin{array}{l}
\delta 
\end{array}\right] \\
= Ax + Bu,
\label{eqn:statespaceeqns}
\end{equation}
where,
\begin{equation*}
\begin{aligned}
a=-\frac{2C_{f}+2C{r}}{m\dot{x}},&b=-\dot{x}-\frac{2C_{f}l_{f}-2C_{r}l_{r}}{m\dot{x}}\\ 
c=-\frac{2l_{f}C_{f}-2l_{r}C_{r}}{I_{z}\dot{x}},               
&d=-\frac{2l_{f}^{2}C_{f}+2l_{r}^{2}C_{r}}{I_{z}\dot{x}}\\
e=\frac{2C_{f}}{m},
&f=\frac{2l_{f}C_{f}}{I_{z}}.\\
\end{aligned}
\end{equation*}
Table~\ref{tab:vehicleParmeter} lists the vehicle parameter used in all MPC optimization problem techniques.

\begin{table}[]
\centering
\setlength{\tabcolsep}{3mm}
\renewcommand{\arraystretch}{1.5}
\setlength{\arrayrulewidth}{0.2mm}
\caption{vehicle model Parmeter}
\label{tab:vehicleParmeter}
\begin{tabular}{|c|c|c|c|}
\hline Parameter            & Symbol         & Value             & Units \\
\hline
Vehicle mass  & ${m}$          & 1,650             & kg \\
\hline Yaw inertia          & ${I}_{z}$      & 2,650     & kg$\cdot$ $\textnormal{m}^{2}$ \\
\hline Front axle to CG     & ${I}_{f}$      & 1.1               & m \\
\hline Rear axle to CG      & ${I}_{r}$      & 1.7               & m \\
\hline Cornering stiffness of front-axle     & $C_f$ & 55,494    & N/rad \\
\hline Cornering stiffness of rear-axle      & $C_{r}$ & 55,494   & N/rad \\
\hline
\end{tabular}
\end{table}


\section{Mathematical model of a free-flying robot}
\label{sec:S.robotMdl}
 \begin{figure}[]
 \centerline{\includegraphics[width=.4\textwidth]{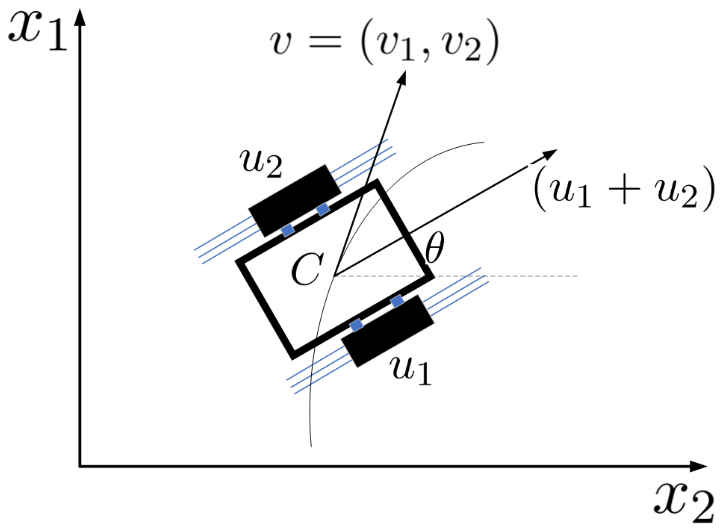}}
 \caption{robot Model}
 \label{fig:robotmdl}
 \end{figure}
A planner motion of a free-flying robot (FFR) shown in Fig. S.2~\cite{sakawa1999robotmdl}, the flying robot has six states ($x_{1},x_{2},\theta,v_{1},v_{2},\omega$) and two thrusters ($u_{1},u_{2}$) to move in 2-D space. ($x_{1}, x_{2}$) are the inertial coordinates of the center of gravity $C$ of the FFR and $\theta$ is the direction of thrust of the FFR. ($v_{1},v_{2}$) are the velocity vector of the FFR at $C$. $\omega$ is the angular velocity of the FFR. $\mathbf{u} = (u_{1},u_{2}) $ is the control vector representing the thrust due to the jet fuel flow.

We assume that the two thrust mechanisms of the FFR are of the same structure and parallel to each other. Then, the equations of motion of the FFR are given by~\cite{sakawa1999robotmdl},
\begin{equation}
    \begin{aligned}
    & \dot{x_{1}} = v_{1} \\
    & \dot{x_{2}} = v_{2} \\
    & \dot{\theta} = \omega \\
    & \dot{v_{1}} = (u_{1} + u_{2}) cos(\theta)\\
    & \dot{v_{2}} = (u_{1} + u_{2}) \sin(\theta)\\
    & \dot{\omega} = \alpha u_{1} - \beta u_{2}, \\
    \end{aligned}
    \label{eqn:robotMdl}
\end{equation}
where  $\alpha = d_{1} m/J$ and $\beta = d_{2} m/J$  where $(d_{1}$ and $d_{2})$ are the distances from $C$ to the jet thrust mechanisms, respectively, $m$ is the total mass of the FFR, and $J$ is the moment of inertia of the FFR at $C$. Each control is constrained by the maximum fuel flow so we assume that~\cite{sakawa1999robotmdl}
\begin{equation*}
\left|u_{i}\right| \leq M \quad(i=1,2),
\end{equation*}
where $M$ is a constant. The FFR equations of motion \eqref{eqn:robotMdl} describe a set of non-linear equations, and the translational motion and the rotational motion are coupled to each other which can be described as 
 
\begin{equation*}
\dot{\mathbf{x}}(t)=\mathbf{f}(\mathbf{x}(t), \mathbf{u}(t))
\end{equation*}
Consequently, a non-linear system in state space form can be written as 
\begin{equation}
\left[\begin{array}{c}
\dot{x_{1}} \\
\dot{x_{2}} \\
\dot{\theta} \\
\dot{v_{1}} \\
\dot{v_{2}} \\
\dot{\omega} 
\end{array}\right]
=\left[\begin{array}{cccccc} 
0 & 0 & 0 & 1 & 0 & 0 \\
0 & 0 & 0 & 0 & 1 & 0 \\
0 & 0 & 0 & 0 & 0 & 1 \\
0 & 0 & 0 & 0 & 0 & 0 \\
0 & 0 & 0 & 0 & 0 & 0 \\
0 & 0 & 0 & 0 & 0 & 0 
\end{array}\right]
\left[\begin{array}{c}
x_{1} \\
x_{2} \\
\theta \\
v_{1} \\
v_{2} \\
\omega 
\end{array}\right]
+\left[\begin{array}{cccccc} 
0 & 0 \\
0 & 0 \\
0 & 0 \\
\cos{\theta} & \cos{\theta} \\
\sin{\theta} & \sin{\theta} \\
\alpha & -\beta 
\end{array}\right]
\left[\begin{array}{c}
u_{1} \\
u_{2} \\
\end{array}\right] \\
 = A(t)x(t) + B(t)u(t) 
\label{eqn:statespaceeqnsRobot}
\end{equation}
In all our experiments we set $\alpha = \beta = 0.2$ and  $M = 1$.

\section{Experimental settings}
\label{sec:S.Expsettings}
\begin{figure}
     	\centerline{\includegraphics[width=.4\textwidth]{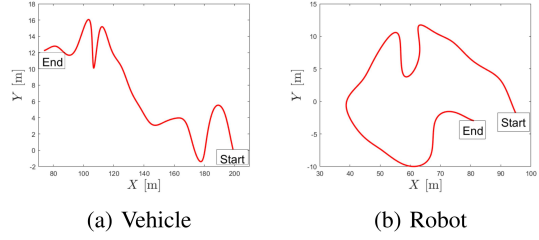}}
         \caption{Vehicle and robot reference spatial trajectories.}
     \label{fig:expSpatialRef}
\end{figure}

\begin{figure}
     \centerline{\includegraphics[width=.9\textwidth]{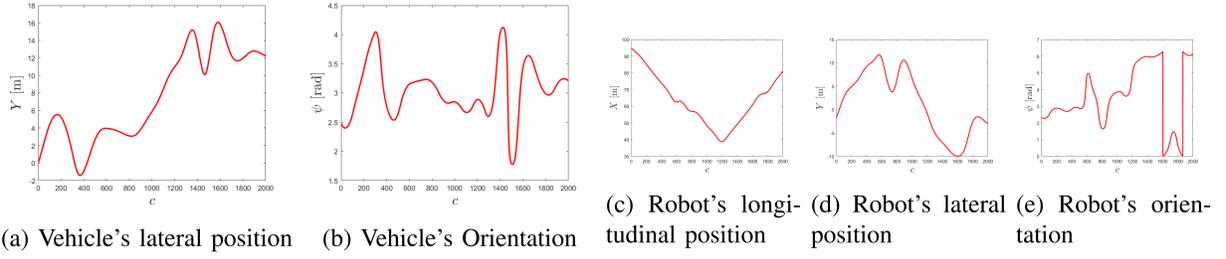}}
        \caption{Reference state trajectories for both the vehicle and robot models. These reference state trajectories are chosen to contain both rabid and slow variation as in reality.}
         \label{fig:expStatesRef}
\end{figure}

Figure~\ref{fig:expSpatialRef} shows the reference spatial trajectories for both the vehicle and the robot for the performed experiments in Sec. 5 in the main manuscript. Figure~\ref{fig:expStatesRef} shows their reference state trajectories, which as in the reality, they may contain both rabid and slow variations as shown in the figure.
The parameters of the MPC's optimization~(7) and the constraints settings in the main manuscript for all compared techniques are kept the same and shown in Table~\ref{tab:MPCAlgorithmparameters} and Table~\ref{tab:MPCAlgorithmconstraine}, respectively. Table~\ref{tab:constrainedMPC} presents the tightly constrained MPC optimization parameter settings used by the second experiment in Sec.5.3 in the main manuscript.

\begin{table}
\renewcommand{\arraystretch}{1.5}
\centering
\captionsetup[table]{position=top}
\caption{MPC optimization function parameter settings}      
\begin{tabular}{|l|l|l|}
\hline
Parameter & Vehicle   & Robot \\ \hline
$Q$  & diag(10,50,10,100)    & diag(20,20,50,10,10,20)  \\ \hline
$R$      & [5]      & [10 10] \\ \hline
$t_{s}$              & .02     & .02      \\ \hline
\end{tabular}
\label{tab:MPCAlgorithmparameters}
\end{table}

\begin{table}
\renewcommand{\arraystretch}{1.5}
\centering
\captionsetup[table]{position=top}
\caption{MPC optimization constraints settings }          
\begin{tabular}{|l|l|l|}
\hline
Parameter & Vehicle   & Robot \\ \hline
$[u_{\min} ,u_{\max}]$ &
$ [\frac{-\pi}{6},\frac{\pi}{6}]$      & [-1,1]     \\ \hline
$[\Delta u_{\text{min}}, \Delta u_{\text{max}}] $ & $ [\frac{-\pi}{12},\frac{\pi}{12}]$      & [-.1,.1]      \\
\hline
$\mathbf{x_{\text{min} } } $ & $[-.5, -inf, \frac{-\pi}{6}, -inf] $      &  $[-inf, -inf, -inf, -.2, -.2, \frac{-\pi}{12}]  $    \\
\hline
$\mathbf{x_{\text{max} } } $ & $[.5, inf, \frac{\pi}{6}, inf] $      &  $[inf, inf, inf, .2, .2, \frac{\pi}{12}]  $    \\
\hline
\end{tabular}
\label{tab:MPCAlgorithmconstraine}
\end{table}

\begin{table}
\renewcommand{\arraystretch}{1.5}
\centering
\captionsetup[table]{position=top}
\caption{Tightly constrained MPC optimization parameter settings }          
\begin{tabular}{|l|l|l|}
\hline
Parameter & Vehicle   & Robot \\ \hline
$[u_{\min} ,u_{\max}]$ &
$ [\frac{-\pi}{6},\frac{\pi}{6}]$      & [-1,1]     \\ \hline
$[\Delta u_{\text{min}}, \Delta u_{\text{max}}] $ & $ [\frac{-\pi}{24},\frac{\pi}{24}]$      & [-.01,.01]      \\
\hline
$\mathbf{x_{\text{min} } } $ & $[-.05, -inf, \frac{-\pi}{24}, -inf] $      &  $[-inf, -inf, -inf, -.02, -.02, \frac{-\pi}{30}]  $    \\
\hline
$\mathbf{x_{\text{max} } } $ & $[.05, inf, \frac{\pi}{24}, inf] $      &  $[inf, inf, inf, .02, .02, \frac{\pi}{30}]  $    \\
\hline
\end{tabular}
\label{tab:constrainedMPC}
\end{table}

In the following, we present the settings for other techniques used in our comparison in the experimental results section (Sec. 5) in the main manuscript.

\subsection{Adaptive dual MPC (ADMPC) settings}
\label{ssec:S.ADMPC}

The adaptive dual MPC technique (ADMPC) divides the horizon length into two parts: a dense part for the near future and a sparse part for the distant future. So, the horizon length $T_c$ in the $c^{\text{th}}$ control cycle is
\begin{equation}
T_{c} = N_{d} + N_{s}n_{s},
\label{eqn:HorizonLength}
\end{equation}
where $N_{d}$ is the number of samples in the dense part, $N_{s}$ is the number of samples in the sparse part, and $n_{s}$ is the relative duration of the sparse samples. At each cycle, ADMPC changes the horizon length by modifying the variable $n_{s}$ in~\eqref{eqn:HorizonLength} according to the reference states trajectory curvature with respect to curvature threshold $C_{\text{th}}$. ADMPC parameters used in both the first and the second experiments in Sec.5.1 in the main manuscript are listed in Table~\ref{tab:ADMPCsettings}.

\begin{table}[]
\caption{ADMPC parameter settings}
\setlength{\tabcolsep}{3mm}
\setlength{\arrayrulewidth}{0.2mm}
\renewcommand{\arraystretch}{1.5}
\center
\begin{tabular}{| m{2cm} | m{1.5cm}| m{1cm} | m{1cm} | m{1cm} |}
\hline
 \multirow{2}{1.5cm}{Parameter} & \multicolumn{2}{c|}{Experiment one}                               &\multicolumn{2}{c|}{Experiment two}\\
&{Vehicle}&{Robot}&{Vehicle}&{Robot} \\ 
\hline
 $N_{d}$  & 6	& 8	& 4	& 6 \\  
   \hline
 $N_{s}$ & 10	& 10 & 8	& 8\\
   \hline
 Minimum $n_{s}$ & 1 &1&1&1\\
  \hline
 Maximum $n_{s}$ &20	&20	&20	&20\\
  \hline 
 $C_{\text{th}}$ & 0.01	& 0.01	& 0.005	& 0.005\\
  \hline
\end{tabular}
\label{tab:ADMPCsettings}
\end{table}

\subsection{Adaptive neural network MPC (ANMPC) settings}
\label{ssec:S.ANMPC}
The horizon length and sample count for the Adaptive neural network MPC (ANMPC) are similar to the ADMPC technique except for the adaptation of the horizon length is performed by modifying the variable $N_{s}$ in~\eqref{eqn:HorizonLength} instead of $n_{s}$. This allows the ANMPC technique to update the overall sample count and consequently update the horizon length. ANMPC parameters used in both the first and the second experiments in Sec.5.1 in the main manuscript are listed in Table~\ref{tab:ANMPCsettings}.

\begin{table}[]
\caption{ANMPC parameter settings}
\setlength{\tabcolsep}{3mm}
\setlength{\arrayrulewidth}{0.2mm}
\renewcommand{\arraystretch}{1.5}
\center
\begin{tabular}{| m{2cm} | m{1.5cm}| m{1cm} | m{1cm} | m{1cm} |}
\hline
 \multirow{2}{1.5cm}{Parameter} & \multicolumn{2}{c|}{Experiment one}                               &\multicolumn{2}{c|}{Experiment two}\\
&{Vehicle}&{Robot}&{Vehicle}&{Robot} \\ 
\hline
 $N_{d}$  & 3	& 3	& 2	& 2 \\  
   \hline
Minimum $N_{s}$ & 1	& 1 & 1	& 1\\
   \hline
Maximum $N_{s}$ &30	&35 &15	&20\\
   \hline
 $n_{s}$ & 4	&4	&4	&4\\
  \hline
\end{tabular}
\label{tab:ANMPCsettings}
\end{table}

\section{Simulation video for controlling the vehicle}
\label{sec:S.video}
\begin{figure}[h!]
        \centerline{\includegraphics[width=.9\textwidth]{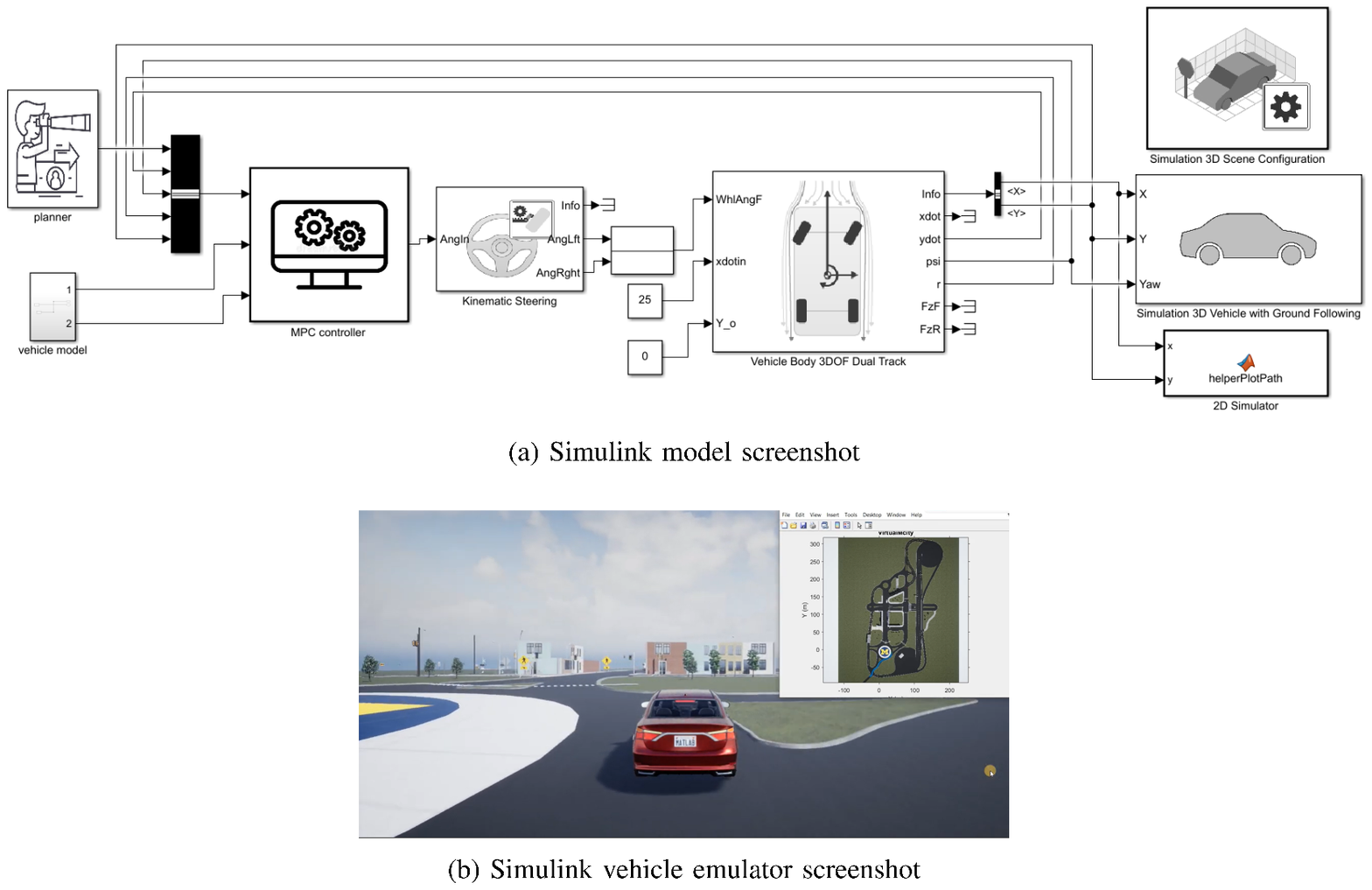}}
        \caption{A simulation for the vehicle controlled by the proposed ARMPC}
         \label{fig:screnshots}
\end{figure}

We simulate our proposed technique using a Matlab/Simulink model and record a video for the whole simulation. Our video is published here~\cite{SimVideoYouTube}. The video shows a simulation of a vehicle controlled by the proposed ARMPC. In this video, we show how we draw a spatial reference trajectory on a real map which is transformed into reference state trajectories to be controlled by the ARMPC. Figure.~\ref{fig:screnshots} shows screenshots of the Simulink model and the vehicle emulator that are used in our simulation.
